\documentclass[twocolumn,superscriptaddress,prb,showpacs,10pt,amsfonts,amsmath,amssymb,nobibnotes]{revtex4-1}
\usepackage[colorlinks=true,urlcolor=blue]{hyperref}
\usepackage{tgtermes}
\usepackage{booktabs}
\usepackage{graphicx}
\usepackage{subfig}
\usepackage{amsmath,amsfonts}
\begin{document}
\title{Persisting of Polar Distortion with Electron Doping in Lone-Pair Driven Ferroelectrics} 
\author{Xu He}
\affiliation{Beijing National Laboratory for Condensed Matter Physics, Institute of Physics, Chinese Academy of Sciences, Beijing 100190, China}
\author{Kui-juan Jin}
\email[Correspondence and requests for materials should be addressed to Kui-juan Jin. Email: ]{kjjin@iphy.ac.cn}
\affiliation{Beijing National Laboratory for Condensed Matter Physics, Institute of Physics, Chinese Academy of Sciences, Beijing 100190, China}
\affiliation{Collaborative Innovation Center of Quantum Matter, Beijing 100190, China}
\begin{abstract}
 Free electrons can screen out long-range Coulomb interaction and destroy the
 polar distortion in some ferroelectric materials, whereas the coexistence of
 polar distortion and metallicity were found in several non-central-symmetric
 metals (NCSMs). Therefore, the mechanisms and designing of NCSMs
 have attracted great interests. In this work, by first-principles calculation,
 we found the polar distortion in the lone-pair driven ferroelectric material
 PbTiO$_3$ can not only persist, but also increase with electron doping. We
 further analyzed the mechanisms of the persisting of the polar distortion. We
 found that the Ti site polar instability is suppressed but the Pb site polar
 instability is intact with the electron doping. The Pb-site instability is due
 to the lone-pair mechanism which can be viewed as a pseudo-Jahn-Teller effect,
 a mix of the ground state and the excited state by ion displacement from the
 central symmetric position. The lone-pair mechanism is not strongly affected by
 the electron doping because neither the ground state nor the excited state
 involved is at the Fermi energy.  The enhancement of the polar distortion is
 related to the increasing of the Ti ion size by doping. These results show that
 the long-pair stereoactive ions can be used for designing NCSMs.
\end{abstract}
\maketitle

\section{Introduction}
In ferroelectric materials, there is a delicate balance between the short-range repulsion which favors the non-polar
structure and the long-range Coulomb interaction which favors the ferroelectric
state\cite{doi:10.1080/00018736000101229,0295-5075-33-9-713}.  Free carriers
screen out the long-range Coulomb interaction, thus it can reduce the polar
distortion. For example, in a prototypical ferroelectric material BaTiO$_3$, the
ferroelectric distortion is weakened with electron doping, and eventually
disappears when a critical concentration is 
reached~\cite{BaTiO3dopingl03,PhysRevLett.101.205502,wang2012ferroelectric,iwazaki2012doping}.
The screening of the Coulomb interaction was believed to be the reason for the
weakening or disappearing of the polar distortion for
BaTiO$_3$~\cite{wang2012ferroelectric,iwazaki2012doping}. Although the screening
effect of the carriers can also inhibit the polar distortion in many other
ferroelectric materials, there are exceptions which have attracted a lot of
attentions and efforts. Anderson and Blount pointed out that while free
electrons can screen out the electric field, the transverse optical soft phonons
can lead to polar distortion~\cite{anderson1965symmetry}. A few ``ferroelectric"
metals, or more precisely non-central-symmetric metals (NCSMs) have been found
or proposed, such as perovskite structure
LiOsO$_3$\cite{shi2013ferroelectric,PhysRevB.89.201107,PhysRevB.90.094108,PhysRevB.90.195113,PhysRevB.91.064104},
MgReO$_3$~\cite{PhysRevB.90.094108}, and the cation-ordered
SrCaRu$_2$O$_6$\cite{puggioni2014designing}. The possible mechanisms of the
NCSMs have been discussed by several authors. Xiang suggested that the long-range
Coulomb interaction is not necessary for the polar distortion; the short-range
pair interactions, which are not screened out by free electrons, can drive the
polar distortion~\cite{PhysRevB.90.094108}. Puggioni and Rondinelli proposed
that the non-centrosymmetric structure can exist if the coupling of the soft
phonon mode and the electrons at the Fermi level is
weak~\cite{puggioni2014designing}. Benedek and Birol proposed that polar
distortion can emerge through a geometric mechanism in metals\cite{C5TC03856A}.

  Although the NCSM structures are no longer suitable for the usage as ferroelectric materials because of the metallicity, other interesting properties are found in NCSMs, like the unconventional optical responses~\cite{PhysRevB.81.094525,PhysRevB.83.113109}, magnetoelectricity~\cite{PhysRevLett.75.2004}, superconductivity~\cite{0953-8984-8-3-012}, and thermoelectricity\cite{puggioni2014designing}. A deeper understanding of the mechanisms of NCSM could help finding new NCSMs. The purpose of this work is to seek a possible mechanism of NCSMs so new NCSMs can be found or designed.

In this work, by studying the polar
distortion in electron-doped PbTiO$_3$, we show that the lone-pair driven polar
distortion is compatible with metallicity. The lone pair mechanism for the polar
distortion in non-doped PbTiO$_3$ has been long
studied~\cite{cohen1992origin,cohen1992electronic,PhysRevLett.87.217601,PhysRevB.74.172105}.
In PbTiO$_3$, the hybridization between the Pb (6s, 6p) bands and O 2p bands 
reduces the short-range repulsion, resulting in a large polar distortion, which
is often referred as lone-pair driven ferroelectricity.  The lone-pair driven
ferroelectricity can be interpreted as the result of the Pseudo-Jahn-Teller effect (PJTE). 
\cite{BERSUKER1966589,doi:10.1080/00150197808237842,bersuker2006,doi:10.1021/cr300279n}
which is local. Therefore, it is possible that the lone pair mechanism can still
drive the polar distortion even if the long-range interaction is screened.
We found that the polar distortion in PbTiO$_3$ not only persists 
but also is enhanced with the electron doping. Then by analyzing the evolution
of the phonon and the force constant matrices (FCM's), we found that the A-site
instability, which is caused by the lone-pair mechanism, is responsible for the persisting of the polar distortion, 
 because the electronic states involved are far away
from the Fermi energy. We showed that enhancement of
the polar distortion is related to the
increasing of the Pb-O distance. We discussed the
generalizability of the results from PbTiO$_3$ to other lone-pair driven
ferroelectric materials and propose that the lone-pair stereoactive ions can be
used for the designing of NCSMs. 
\section{Methods}
The density functional theory (DFT) calculations were carried out by using the projected augmented wave (PAW)~\cite{kresse1999ultrasoft} pseudopotentials as implemented in the Vienna \textit{ab initio} simulation package (VASP)~\cite{kresse1996efficient}. The energy cutoff of the plane wave basis set was 500 eV. The exchange-correlation functional with local density approximation (LDA) as parameterized by Perdew and Zunger (PZ)~\cite{perdew1981self} was used. The reference electronic configurations for the pseudopotentials are Pb $5d^{10}6s^{2}6p^2$, Ba $5s^25p^66s^2$, Ti $3s^23p^63d^4$, and O $2s^22p^6$, respectively. An $8\times8\times 8$ $\Gamma$-centered $k$-point mesh was used to represent the reciprocal space. The calculated $P4mm$ structure has a tetragonality ($c/a$) of 1.04 and a volume of 60.2 \AA$^3$, agreeing well with the experimental data and previous DFT calculations~\cite{Shirane:a01619,PhysRevLett.80.4321}.

Electron doping is added into the structure with additional neutralizing background charges, instead of dopants like oxygen vacancies. In this case, however, the energy converges very slowly with respect to the size of the supercell. The problem is solved by adding a first order correction (image charge correction) to the total energy. 
The phonon frequencies and the FCM's were calculated using the density
 functional perturbation theory\cite{phonondfpt} (DFPT) method as implemented in
 VASP and the phonopy\cite{phonopy} package. The non-analytic
 contribution (NAC) \cite{PhysRevB.55.10355} to the phonon and
 FCMs\cite{PhysRevB.1.910} was not considered. The Born effective charges (BEC's) for
 structures with free charges are not well defined, inhibiting the calculation
 of the NAC. The NAC only affects the longitudinal optical (LO) phonon frequencies,
 whereas the polar distortion is caused by the transverse optical (TO) phonon. We
 also checked the phonon results with NAC added against others'
 work\cite{PhysRevLett.72.3618} and the results agree well. The comparisons of BEC's and the
 $\Gamma$-point phonon frequencies for the $Pm\bar{3}m$ structure are in
 table~\ref{tab:BEC} and table~\ref{tab:phonon}, respectively.The $\Gamma$ point FCMs were calculated in a
 $1\times 1\times 1$ $Pm\bar{3}m$ structure. A $2\times 2\times 2$ supercell of
 40 atoms was used for the phonon band calculation.\footnote{Although DFPT is
   capable of calculating the dynamic matrix at arbitrary wave vector without
   using supercells, only the $\Gamma$-point result is outputted in the VASP
   implementation.  Phonopy makes use of a larger supercell to calculate the
   phonon frequencies at a few other wave vectors and then make a 
   Fourier interpolation to calculate the full phonon dispersion curve.} The
 phonon frequencies were calculated with the structures fully relaxed. A
 $\Gamma$-centered $20\times20\times20$ Monkhosrt-Pack grid was used to
 calculate the phonon density of states (PDOS). The crystal orbital hamilton
 population~\cite{COHP,pwCOHP} (COHP) analyses were carried out with the
 LOBSTER~\cite{LOBSTER} code. The crystal structures and the contour maps of the
 electron localization functions~\cite{elfcar} (ELFs) were visualized using the
 VESTA package~\cite{vesta}. 

The ferroelectric polarization in a structure with free charges is not well
defined. Therefore, the structural polar distortion is discussed instead of the
ferroelectric polarization in the present work.
\section{Results and Discussion}
\begin{figure}[htbp]
  \centering
  \includegraphics[width=0.38\textwidth]{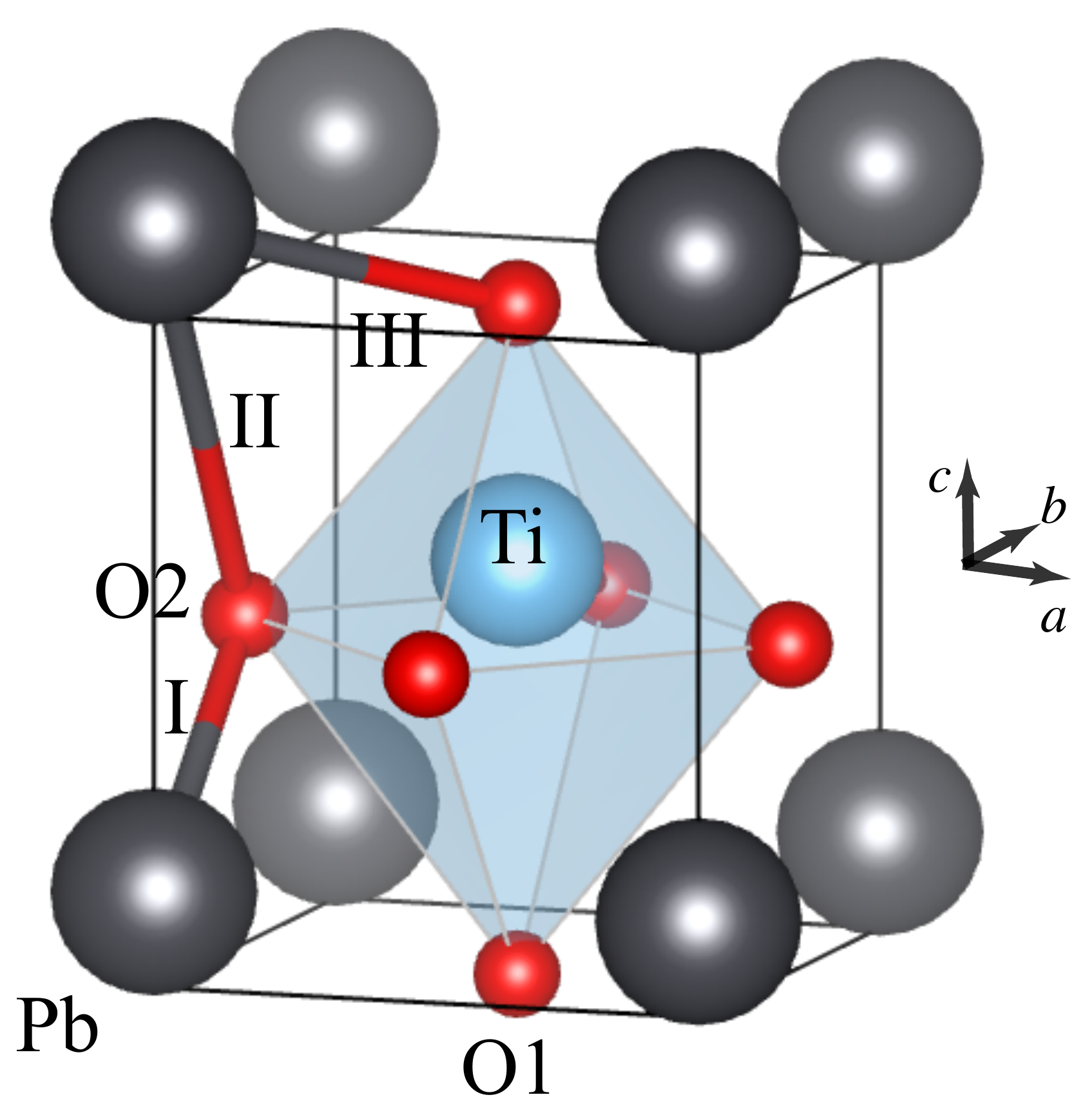} 
  \caption{The structure of $P4mm$ PbTiO$_3$. The black, blue, and red spheres
    are the Pb, Ti, and O ions, respectively. O1 and O2 are the in-plane and
    apical oxygen ions, respectively. Due to the displacements of the ions in the
    $c$ direction, there are two kinds of Pb-O2 bonds, the shorter one (type I)
    and the longer one (type II). The Pb-O1 bonds are labeled as type III. }
  \label{fig:structPTO}
\end{figure}

\begin{figure*}[htpb]
  \centering
  \includegraphics[width=0.84\textwidth]{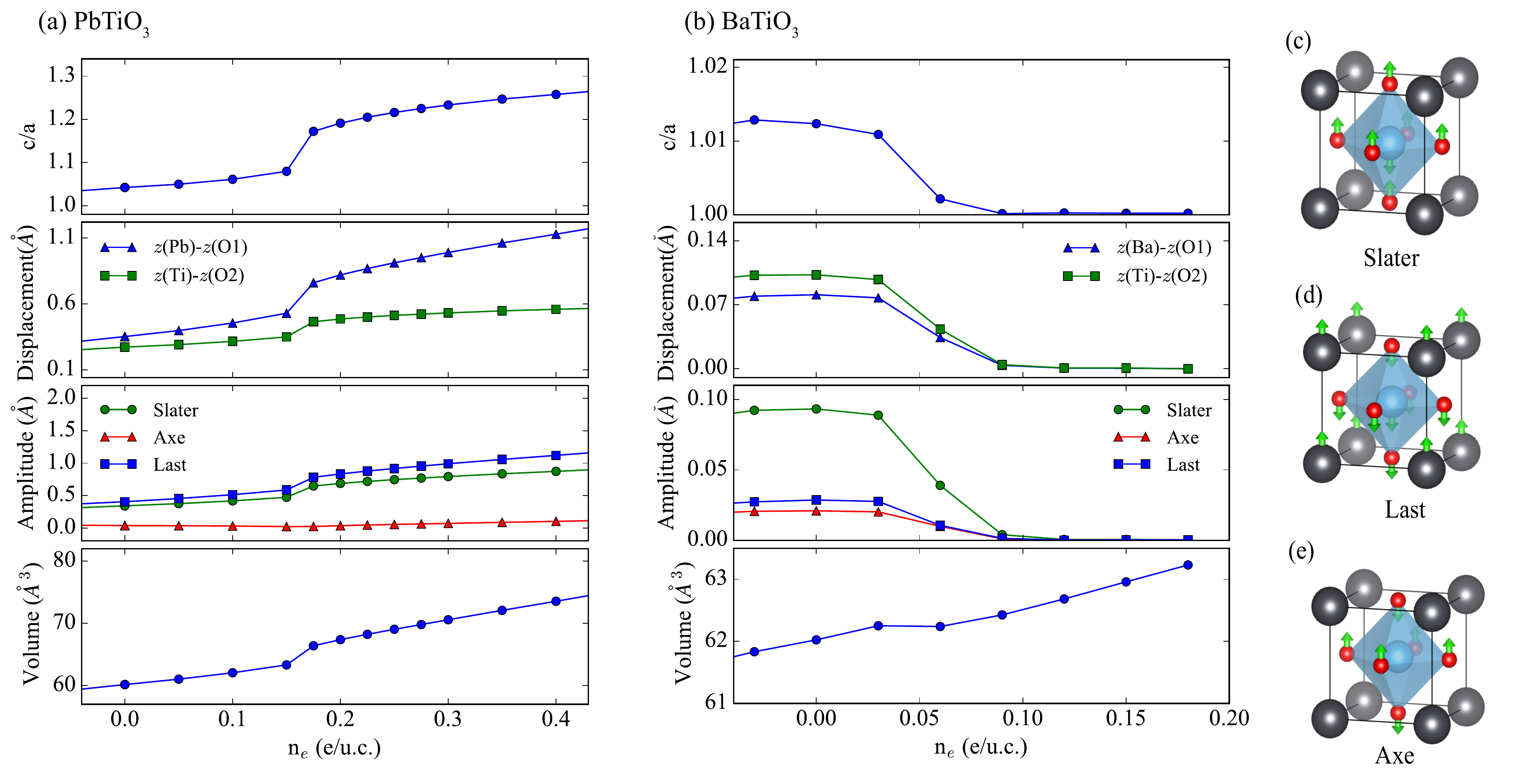}
  \caption{ The tetragonality (ratio of out-of-plane lattice constant $c$ and
    the in-plane lattice constant $a$), the relative displacements of Pb-O ($z(Pb/Ba)-z(O1)$) and
    Ti-O ($z(Ti)-z(O2)$), the amplitudes of the polar modes, and the volume as functions of doped electron density $n_e$ in (a)
    PbTiO$_3$ and BaTiO$_3$ are shown in (a) and (b), respectively. (c), (d),
    and (e) are the polar normal modes, namely the Slater, Axe, and Last modes.}
  \label{fig:structvare}
\end{figure*}

Here we show the enhancement of the polar distortion with electron doping in
PbTiO$_3$. The ferroelectric PbTiO$_3$ structure with $P4mm$ symmetry group is
shown in Fig.~\ref{fig:structPTO}, where the Pb and Ti cations displace along the
$c$ direction, and the O anions displace in the opposite direction.
The change of the polar distortion in $P4mm$ PbTiO$_3$ structure with the concentration
of the doped carriers ($n_e$) is shown in Fig.~\ref{fig:structvare} (a). As $n_e$
increases, the tetragonality ($c/a$) of the lattice and the relative
displacements of the Pb and Ti cations from the O anions increase. These
phenomena show the enhancement of the polar distortion, in contrast to what happens in the $P4mm$
BaTiO$_3$\cite{BaTiO3dopingl03,PhysRevLett.101.205502,wang2012ferroelectric,iwazaki2012doping},
where the tetragonity and the cation-anion relative displacements both decreases to
zero when the electron doping level is about 0.1 $e$/u.c., as shown in
Fig.~\ref{fig:structvare} (b). The polar distortion can be decomposed into three
$\Gamma$-point normal modes\cite{PhysRevB.84.104440} (Fig.~\ref{fig:structvare}), namely the Slater,
Last, and Axe modes. The Slater mode\cite{slatermode}
(Fig.~\ref{fig:structvare} (c)) involves the displacement of the Ti ion from the
center of the oxygen octahedron, which is corresponding to the Ti-site
instability. The Last mode\cite{lastmode} (Fig.~\ref{fig:structvare} (d)) involves
the displacement of the A-site ions against the TiO$_6$ octahedron, which is
corresponding to the A-site instability. The Axe mode\cite{axemode}
(Fig.~\ref{fig:structvare} (e)) involves the relative displacement of the O1 and O2
ions, which is corresponding to the distortion of the oxygen octahedron. It can be seen that in
PbTiO$_3$, both Slater mode and Last mode contribute to the polar distortion.
The amplitudes of them both increase as electrons are doped. For BaTiO$_3$,
Slater mode is predominant and decreases with electron doping. The difference
implies that the Pb-site instability may drive the enhancement of the polar
distortion in PbTiO$_3$. 

We also found that the lattice volumes increase with $n_e$ in both PbTiO$_3$ and BaTiO$_3$ (Fig.~\ref{fig:structvare}). 

In our calculation, the doping level ranges from 0 to about 0.4 e/u.c.. Such
high doping level is hard or even impossible to be realized by adding dopants
such as oxygen vacancies to the structure without changing the structural
properties dramatically. Therefore, the results for high-doping levels presented here may not be
applicable directly. However, it is still beneficial to see what happens in
the hypothetical heavily doped structure in which the long-range  Coulomb interaction
should be mostly screened out. 

In BaTiO$_3$, the decreasing of the polar distortion with electron doping was
believed to be because of the screening effect~\cite{wang2012ferroelectric,iwazaki2012doping}. The doped electrons screen out the long-range Coulomb interaction. The screening length $\lambda$ can be estimated with the Thomas-Fermi model $\lambda=\sqrt{\varepsilon/e^2 D(E_F)}$, where $\varepsilon$ is the dielectric permittivity, and $D(E_F)$ is the density of states (DOS) at the Fermi level. As the concentration of doped electrons increases, the DOS at the Fermi energy increases, and thus the screening length decreases. The values of $\varepsilon$ and the values of $D(E_F)$ for PbTiO$_3$ are close to those for BaTiO$_3$, respectively. ( $\varepsilon$(PbTiO$_3$) and $\varepsilon$(BaTiO$_3$) are 81$\varepsilon_0$ and 78$\varepsilon_0$ from the LDA and DFPT\cite{PhysRevB.72.035105} calculation\cite{1347-4065-50-9S2-09NE02}, respectively. When the concentration of electrons is 0.2 $e$/u.c., the $D(E_F)$ are 0.7 and 1.1 states/(eV$\cdot$u.c.), and consequently the screening length are about 5 \AA~and 6 \AA~in PbTiO$_3$ and BaTiO$_3$, respectively.) Therefore, the screening effects of the long-range Coulomb interaction are comparable in these two materials. The polar distortion is gradually destroyed by the screening effect in BaTiO$_3$. Then what is the intrinsic reason for the opposite trend of the polar distortion with doped electrons in PbTiO$_3$ from that in BaTiO$_3$?

\begin{figure*}[htbp]
  \centering
  \includegraphics[width=0.95\textwidth]{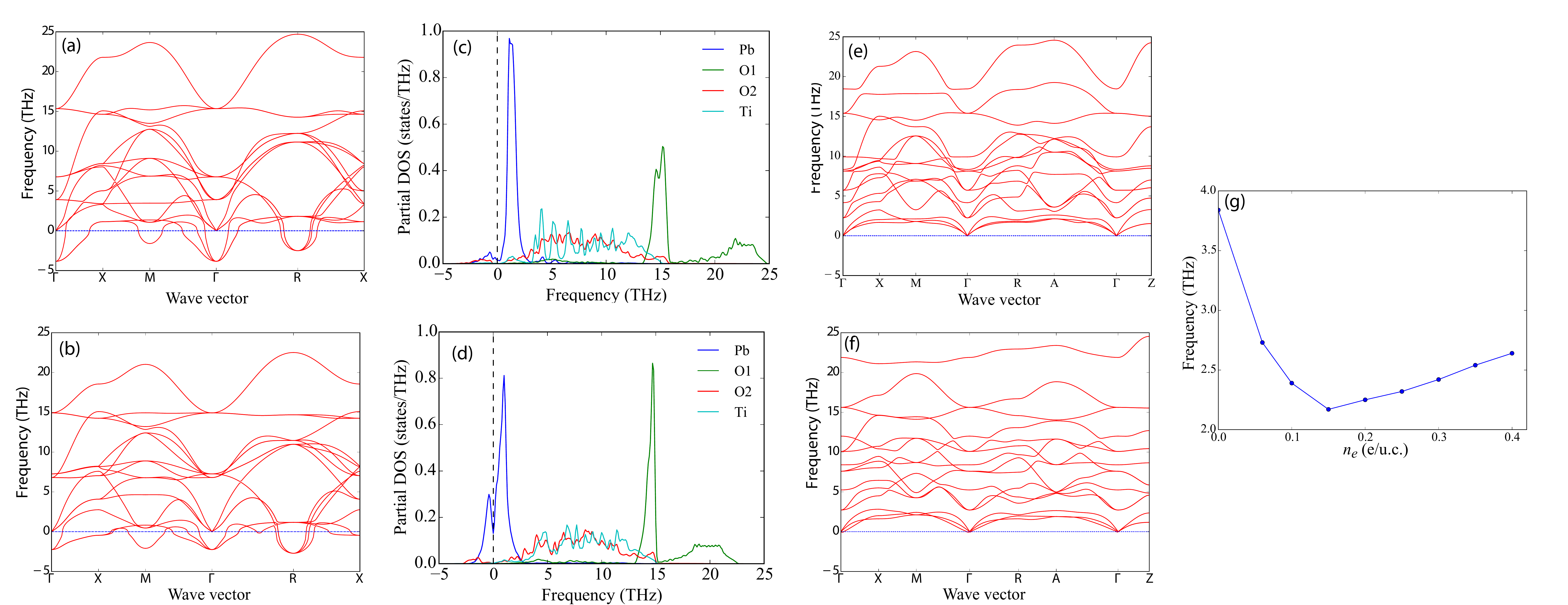}
  \caption{ The phonon band structures of the (a) $Pm\bar{3}m$ structure without
    doping, (b) $Pm\bar{3}m$ structure with electron doping ($n_e=0.2$ e/u.c.),
    (e) $P4mm$ structure without doping, (f) $P4mm$ structure with electron
    doping ($n_e=0.2$ e/u.c.). 
    (c) and (d) are the atom projected partial phonon density of states
    projected on the [001] direction for the $Pm\bar{3}m$ structure without
    doping and with electron doping ($n_e=0.2$ e/u.c.), respectively.
    In (a),
   (b), (c), and (d), the imaginary phonon frequencies are shown as negative
    numbers. The high symmetry q-vectors for $Pm\bar{3}m$ are  $\Gamma$ (0, 0,
    0), X (0, 1/2, 0), M (1/2, 1/2, 0), R (1/2, 1/2, 1/2). The high symmetry
    q-vectors for $P4mm$ are $\Gamma$ (0, 0, 0), X (0, 1/2, 0), M (1/2, 1/2, 0),
    R (0, 1/2, 1/2), A (1/2, 1/2, 1/2), Z (0, 0, 1/2). (g) is the absolute values of the imaginary frequencies at $\Gamma$  as a function of the
    electron doping levels.  \label{fig:phonon}}
\end{figure*}


\begin{table}[htbp]
  \begin{tabular}{l c c c c}
    \hline
    & $Z^*_{Pb}$ & $Z^*_{Ti}$ & $Z^*_{O1,xx}$ & $Z^*_{O1,zz}$ \\
    \hline
    this work & 3.92 & 7.25 & -2.62 & -5.94\\
    Ref.~\onlinecite{PhysRevLett.72.3618} & 3.90 & 7.06 & -2.56 & -5.83\\
    \hline
  \end{tabular}
  \caption{ Born effective charge of PbTiO$_3$ in cubic structure. The unit is 
    $|e|$. Only the $xx$, $yy$, and $zz$ part of the BEC tensors are non-zero.
    For Pb and Ti, $Z^*_{xx}$=$Z^*_{yy}$=$Z^*_{zz}$. For O,
    $Z^*_{O1,yy}=Z^*_{O1,xx}=Z^*_{O2,yy}=Z^*_{O2,zz}$, $Z^*_{O2,xx}=Z^*_{O1,zz}$,  The results in
    Ref.~\onlinecite{PhysRevLett.72.3618} were calculated with LDA. \label{tab:BEC}  }
  \end{table}

  \begin{table}
  \begin{tabular}{l c c c c c c}
    \hline
& TO1 & TO2 & TO3 & LO1 & LO2 & LO3\\
      \hline
    This work & 3.84$i$ & 3.90 &15.32 & 3.15 &12.40 &20.96\\
    Ref.~\onlinecite{PhysRevLett.72.3618}& 4.32$i$ & 3.63 & 14.90 & 3.12 &12.30 &20.18 \\
     \bottomrule\hline
  \end{tabular}
  \caption{$\Gamma$-point phonon frequencies of PbTiO$_3$ in cubic structure
    with NAC added.
    The unit is THz .The results in Ref.
    ~\onlinecite{PhysRevLett.72.3618} were calculated with LDA/DFPT.\label{tab:phonon}}
\end{table}

Aside from the long-range Coulomb interaction, the covalence between Ti and O
causes the Ti-site polar instability in both BaTiO$_3$ and PbTiO$_3$. The
difference is that the covalence between Pb-O also causes the Pb-site polar
instability. Bersuker \textit{et al.} interpreted both the Ti-site and Pb-site
instabilities as the results of PJTE\cite{BERSUKER1966589,doi:10.1080/00150197808237842}.
To see which site is responsible for the polar distortion and how the polar
instabilities are affected by the electron doping, we consider the evolution of the
phonon modes in the paraelectric $Pm\bar{3}m$ phase with the electron doping. The transition to a
ferroelectric phase from a paraelectric phase is featured with imaginary phonon
frequencies in the paraelectric phase. The phonon bands of $Pm\bar{3}m$
phase were calculated, as shown in Fig.~\ref{fig:phonon}.
Without electron doping, the imaginary frequencies are at $\Gamma$, $R$, and $M$
in the $Pm\bar{3}m$ phase, as shown in Fig.~\ref{fig:phonon} (a). With electrons
doped, the imaginary frequency at $\Gamma$ point (Fig.~\ref{fig:phonon} (b))
corresponding to the ferroelectric distortion firstly decreases and then
increases, as shown in Fig.~\ref{fig:phonon} (g), unlike that in
BaTiO$_3$\cite{wang2012ferroelectric}, where the imaginary frequencies disappear
as the concentration of electrons increases. The atom projected phonon densities
of states (PDOS) for $n_e$=0 and $n_e$=0.2 e/u.c. are shown in
Fig.~\ref{fig:phonon} (c) and (d), respectively. For both  $n_e$=0 and $n_e$=0.2
e/u.c., most of the PDOS in the imaginary frequency region is projected on the
Pb sites, and a small portion is projected on the apical Oxygen site O2,
indicating that the soft phonons are from the Pb atoms and the O atoms in the
side plane. The results consist with that the lone
-pair is from the Pb-O electron hybridization~\cite{PhysRevB.91.035112}. 
 \begin{figure*}[htbp]
   \centering
   \includegraphics[width=0.9\textwidth]{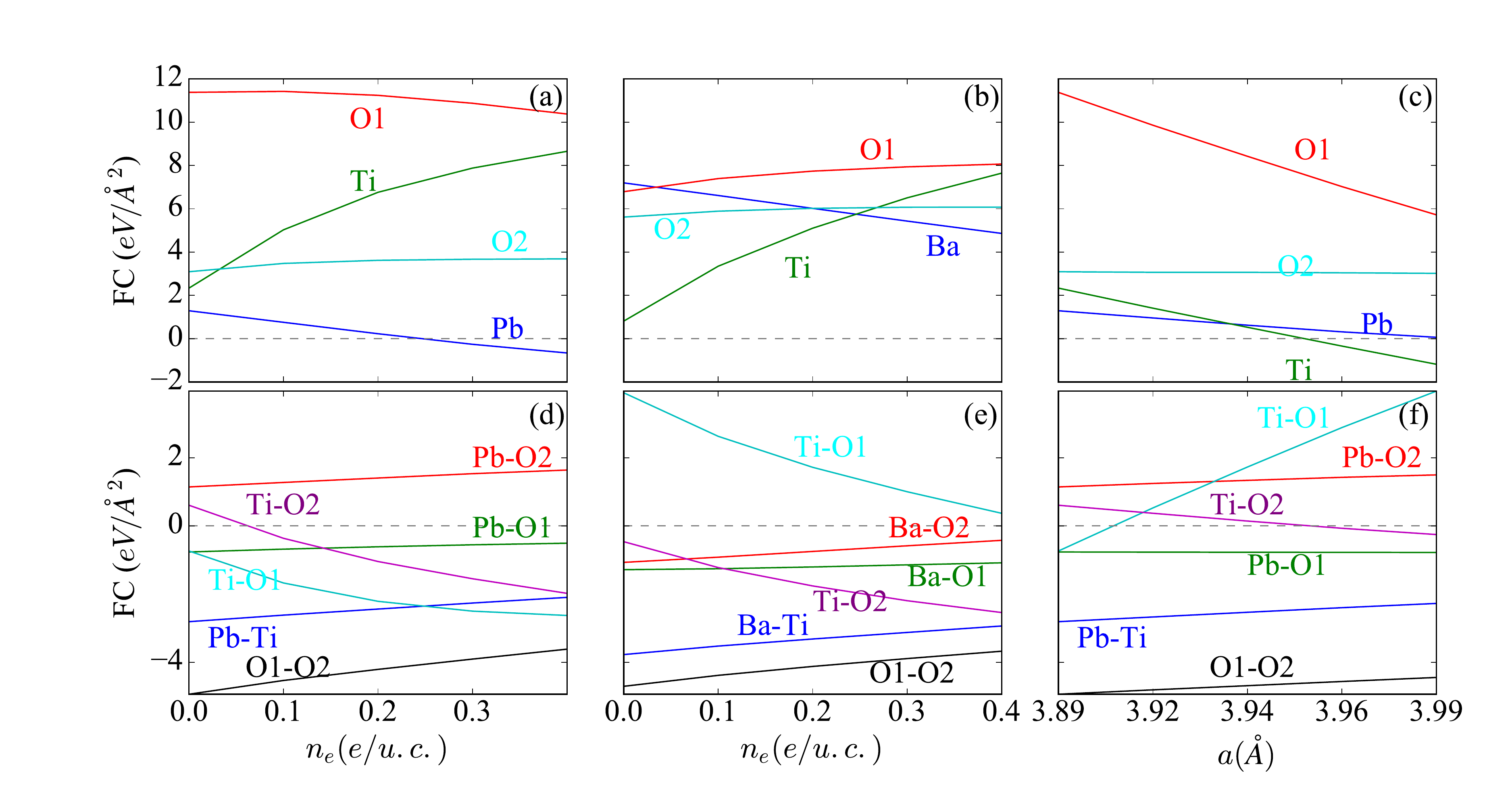}
   \caption{Dependences of the FCM's on the concentration of
     doped electrons. (a) and (b) are the self-force constants on the different
     atoms in $Pm\bar{3}m$ PbTiO$_3$ and BaTiO$_3$, respectively. (d) and (e) are the
     IFC's between different pairs of atoms in PbTiO$_3$
     and BaTiO$_3$, respectively. (c) and (f) are the self-force constants and
     the IFC's in the non-doped $Pm\bar{3}m$ PbTiO$_3$ structures,
     the lattice constants of which were fixed to these with electron doping.
     The lattice constants 
     are corresponing to $n_{el}$ = 0.0-0.4 e/u.c. .}
   \label{fig:FCM}
 \end{figure*}

 We also checked the phonon bands of the $P4mm$ phase with and without electron
 doping, and we found that there are no imaginary frequencies in the phonon
 bands in the $P4mm$ phase (Fig.~\ref{fig:phonon} (e and f)), indicating that the $P4mm$ phase are stable.

To further investigate the change of the lattice ferroelectric instabilities, we calculated the $\Gamma$ point FCM's\cite{PhysRevB.60.836} (FCMs) of the PbTiO$_3$ and BaTiO$_3$ cubic structures with various concentration of doped electrons. The elements of the FCM $D^{xy}_{ij}$=$\partial^2E/\partial u_i\partial u_j$ is the derivative of the energy with the displacements of two atoms $u_i$ and $u_j$, where $u_i$ is along $x$ direction and $u_j$ is along the $y$ direction. Due to the $Pm\bar{3}m$ symmetry, only the $xx$, $yy$, and $zz$ components are none zero. For the same $i$ and $j$ pair, the $xx$, $yy$, and $zz$ components are equal to each other. So we only have to discuss the $zz$ components here, i.e., the displacements are all along the $z$ direction. Thus we omit the superscript of $D^{xy}_{ij}$. The changes of the FCMs are shown in Fig.~\ref{fig:FCM}.

 For $j$ equals to $i$, the element $\partial^2E/\partial u_i^2$, which is the
 second derivative of the energy with the position of the atom, is known as
 self-force constants. A positive (negative) value of the self-force constant
 for an atom indicates an increasing (decreasing) of the energy by solely
 displacing that atom in the supercell with other atoms frozen. The self-force constants of PbTiO$_3$
 and BaTiO$_3$ are shown in Figs.~\ref{fig:FCM} (a) and (b), respectively. In
 both PbTiO$_3$ and BaTiO$_3$, the self-force constants of Ti increase as $n_e$
 increases, which means that if the Ti atoms are displaced from their central
 symmetry positions, the energy costs would be higher, which reduces the Ti-site
 instability. In PbTiO$_3$, the self-force constant of Pb, which is the smallest among all self-force constants, decrease with $n_e$ and eventually get below zero, meaning that the tendency of Pb ions displacing from the central symmetry positions increases. In BaTiO$_3$, though the self-force constant of Ba ion also decreases with $n_e$, it is still much larger than zero, which stabilizes the Ba ion at the central symmetry position.  

Then we look into the interatomic FCs (IFC's).  The IFC's of PbTiO$_3$ and
BaTiO$_3$ are shown in Figs.\ref{fig:FCM} (d) and (e), respectively. A positive
$D_{ij}$ means that the energy would be lowered if the displacements of the two
atoms labeled as $i$ and $j$ are along the different direction. It must be noted
that an IFC is not corresponding to an individual bond but to the
sum of all the interaction between two atoms, which may involve several bonds in
a unit cell due to the periodic boundary condition. For example, if we consider only the nearest neighbor interaction, the Pb-O2 IFC
is corresponding to the bonds of a Pb ion with all the eight nearest O2 ions,
involving 4 type I Pb-O2 bonds and 4 type II Pb-O2 bonds. In PbTiO$_3$ and BaTiO$_3$, the polar distortions are featured with the
anti-parallel displacements of the cations (Pb, Ba, and Ti) and the anions (O),
i.e. the positive values of cation-anion FC constants favor the polar
distortion. In BaTiO$_3$, the antiparallel displacement of Ti and O1 is favored
because of the Ti $3d$-O1 $2p$ hybridization, therefore the IFC is positive. As
$n_e$ increases, the Ti-O1 inter-atomic FC decreases. In PbTiO$_3$, The Ti-O1
and Ti-O2 inter-atomic FC also decreases. The decreasing of Ti-O 
IFC's and the increasing of the Ti self-force constants indicates that the Ti-site
instability is reduced. The Pb-O and Ba-O IFC's increase with $n_e$ in
PbTiO$_3$ and BaTiO$_3$. But only the Pb-O2 IFC is above zero, which drives the
Pb ions away from their central symmetry positions. Whereas the Ba-O interaction
cannot drive the polar distortion. The increasing of A-O FCs and the decreasing
of the A-site self-force constants are the reasons for increasing of the A-site instability. 
%

\begin{figure*}[htbp]
  \centering
  \includegraphics[width=0.75\textwidth]{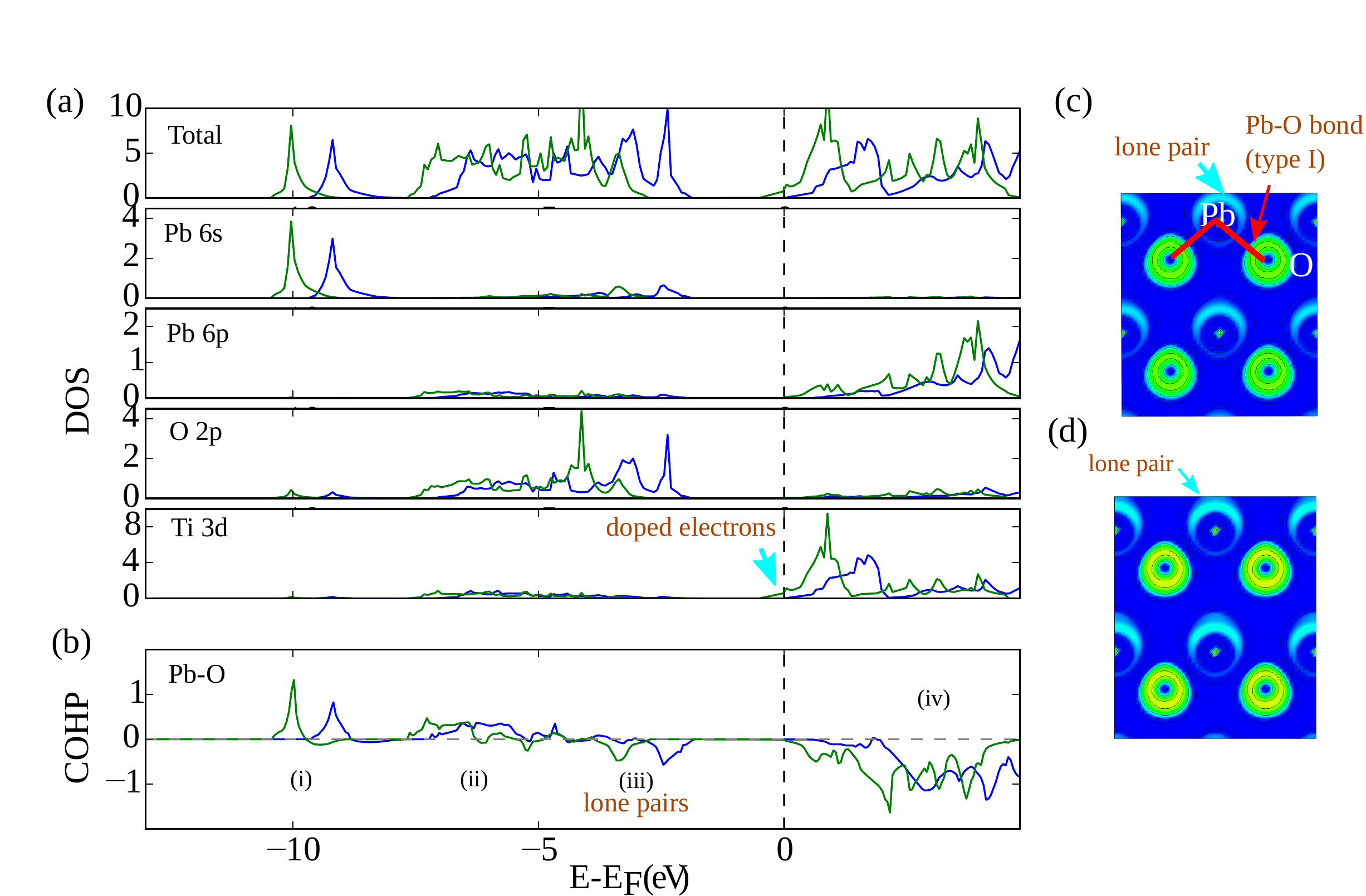}
  \caption{(a) The density of states in PbTiO$_3$. (b) The COHP of the Pb-O bond. In (a) and (b), the blue lines and the green lines represent the results in the non-doped and doped PbTiO$_3$ structures, respectively. The results for non-doped structure are shifted so that the Fermi energy is at the conduction band minimum so that the DOSes for the doped and non-doped structures can be more easily compared. The concentration of the doped electrons is $n_e$=0.2 e/u.c.. (c) and (d) are the contour maps of the ELFs in the $a-c$ PbO plane of the doped and non-doped structures, respectively.}
  \label{fig:lonepair}
\end{figure*}


Both the Ti-site Pb-site instabilities can be viewed as results of PJTE's which
will be referred to as Ti PJTE and Pb PJTE in this work, respectively hereafter. To see
why their responses to the electron doping are different, we calculated the
electronic structure of PbTiO$_3$.

PJTE is through a mix of the ground state and the low excited states by
vibronic coupling. Detailed descriptions of PJTE are in the works of Bersuker
\textit{et al.}\cite{BERSUKER1966589,doi:10.1080/00150197808237842,bersuker2006}. In a high-symmetry $Pm\bar{3}m$ reference system, the energy $E$ can be written as a function of a
normal displacement $q$, where $\partial E/\partial q = 0$. If the curvature $K$ of the energy $E$,
\begin{equation}
  \label{eq:negphon}
  K=(\partial^2E/\partial q^2)_0 
\end{equation}
is negative, the energy is at a local maximum, indicating that the system is unstable.
$E$ is the ground state eigenvalue of Hamiltonian $H$, ($E=\langle \psi_0 | H |
\psi_0 \rangle$). $K$ can be written in two parts, 
\begin{equation}
  \label{eq:expandK}
  \begin{aligned}
    K &= K_0 + K_v \\
    &= \langle \psi_0|(\partial^2 H/ \partial q^2)_0|\psi_0\rangle + 2 \langle \psi_0|(\partial H/ \partial q)_0|\psi_0'\rangle
  \end{aligned}
\end{equation}
, in which, $\psi_0$ is the ground state, $\psi_0'=(\partial \psi_0 / \partial
q)_0$. 

Using the second order perturbation theory, we can get
\begin{equation}
  \label{eq:PJTE}
  K_v=-2\sum_n\frac{|\langle\psi_0|(\partial H/\partial q)_0|\psi_n\rangle|^2}{E_n-E_0}
\end{equation}
, in which $\psi_n$ is an excited state; $E_n$ and $E_0$ are the energies of the
excited state and the ground state, respectively.
It can be noticed that $K_v < 0$  if the mix of the ground state and the excited
state under the displacement $q$ is allowed by symmetry. Thus $K_v$ contributes to the instability. 
If the parity of the product of the ground state and the excited state is odd,
the net overlap of them would be zero in the highest symmetry. However, with a polar
vibration, the hybridization becomes non-zero, causing an energy gain. Therefore,
the BJTE can be interpreted as added covalence in terms of bonding\cite{doi:10.1021/cr300279n}.

The instability of Ti atoms is due to the
PJTE\cite{BERSUKER1966589,doi:10.1080/00150197808237842,bersuker2006}. In the
TiO$_6$ octahedron with Ti $d^0$ electronic configuration, the highest occupied states are the O $2p$
states, with configuration $(t_{1u}\downarrow)^3(t_{1u}\uparrow)^3$, and one
electron transfer to the Ti $3d$ orbitals to form the lowest Excited states with
configuration $(t_{1u}\downarrow)^3(t_{1u}\uparrow)^2(t_{2g}\uparrow)^1$. The net
overlap of them would be canceled out if the Ti cation stays at the center of
the octahedron; the polar mode vibration of the Ti cation
would allow for an overlap and reduce the total energy, leading to a non-zero
$K_v$, which is a driving force of the ferroelectricity. The ``$d^0ness$" plays
an essential role in this kind of PJTE. Electron doping pushes the Fermi energy
into the bottom of the conduction band, which is mostly Ti $3d$
(Fig.~\ref{fig:lonepair} (a)), introducing
$d^1$ electronic configuration, which has no PJTE\cite{PJTEd0d10}. For the $d^1$
configuration, the ground state would be
$(t_{1u}\downarrow)^3(t_{1u}\uparrow)^3(t_{2g}\uparrow)^1$; and the lowest excited state would be
$(t_{1u}\downarrow)^2(t_{1u}\uparrow)^3(t_{2g}\uparrow)^2$. The ground state and
the excited state are of different spin multiplicity, therefore do not mix by
the vibronic coupling. Thus, the Ti PJTE is suppressed as the electron doping increases. 

The lone-pair mechanism of the ferroelectric materials with cations of $s^2p^0$
electronic configuration is another kind of PJTE\cite{doi:10.1080/00150199508221831,lonepairPJTE}. Regarding the Pb-O bonds, the
electronic states near the Fermi-energy are the Pb $6s$ and $6p$ states, and the
O $2p$ states . The occupied Pb $6s$ orbitals and the O $2p$ orbitals form
bonding states with energies of about -10 eV (region (i) in Fig.~\ref{fig:lonepair} (b)), and the anti-bonding states just below the Fermi
energy(region (iii) in Fig.~\ref{fig:lonepair} (b)). These occupied (Pb 6s)-(O 2p) anti-bonding states can be seen as the ground states in
equation ~\eqref{eq:PJTE}; whereas the unoccupied Pb $6p$ states are the
unoccupied excited states. The overlap of them has both positive and negative parts,
which cancels out in the central symmetric cubic structure. However, the coupling between the
ground and excited states becomes non-zero if Pb moves away from the
central-symmetric position, which lowers the total energy and thus drives the
polar distortion. The hybridization between the (Pb 6s-O 2p) bonding states with
Pb $6p$ states results in bonding and antibonding states, corresponding to the
region (ii) and region (iv), respectively. Therefore, the Pb PJTE effect can
also be interpreted as added covalence\cite{doi:10.1021/cr300279n} in the terms
of bonding, which is believed to the driving force
for the Pb-site instability\cite{cohen1992origin,PhysRevB.60.836}. The
displacement of Pb reduces the type I Pb-O bond, leading to a strong covalency
between Pb and O2 ions. The hybridization also causes the asymmetric ELF lobes near at the Pb sites
as shown in Fig.~\ref{fig:lonepair} (c) and (d), which is a characteristic of
the lone-pair mechanism\cite{PhysRevB.91.035112,doi:10.1021/cm010090m}.

As can be seen from Fig.~\ref{fig:lonepair} (a) and (b), the unoccupied Pb $6p$ states,
which are the excited states involved in the Pb PJTE, are above the Fermi
energy when the electrons are doped. Therefore, the Pb PJTE will not be strongly affected by the
electron doping. As a result, the change of the Pb self-force constant with electron
doping is much smaller than that of Ti; the change of the Pb-O inter-atomic
force constants are also much smaller than those of Ti-O, as shown in
Fig.~\ref{fig:FCM}.  There's no significant reduction in the asymmetric lobe of the ELFs in the doped
structure (Fig.~\ref{fig:lonepair} (d))  than that in the non-doped structure (Fig.~\ref{fig:lonepair} (c)).

The increasing of the Pb site instability is likely to be
the result of the increasing of the Pb-O distance with electron doping. The elongation of the Pb-O distance might 
cause a more under-bonded Pb which requires a larger displacement.
The forces between the doped electrons on the Ti $3d$ bands and the negatively
charged O anions are repulsive, which increases the Ti-O bond lengths and the
Pb-O distances are also increased. Thus the overlap of the Pb and O orbitals decreases, which will lead to the decreasing of
both $|K_0|$ and $|K_v|$. If $|K_0|$ decreases more fastly than $|K_v|$, the
instability will increase, for example, Bersuker \textit{et al.} showed that 
$|K_0|$ decreases more fastly than $|K_v|$ in the PJTE involving Ti-site in
RTiO$_3$ (R=Ba, Sr, Ca). In terms of bonding, the PJTE can be interpreted as
added covalent interaction. In perovskites,  a tolerance factor $t$ smaller
(larger) than 1 implies an under-bonded A (B) site ion\cite{PhysRevB.72.054114}. The PJTE is effective
when the added covalent interaction reduces the total energy, which tends to
happen for an under-bonded ion. The increasing of B-site ion size decrease $t$,
thus A-site ion becomes more under-bonded, which tends to enhance the PJTE. To
test whether the enhancement of A-site instability is due to the increased Pb-O
distance, we calculated the IFC's of PbTiO$_3$ without electron doping, and
we fixed the lattice constants of the structures to those with electron doping. The
results are in Fig.~\ref{fig:FCM} (c) and (f). Indeed, the self-force constants
of Pb, the IFC's of Pb-O1 and Pb-O2 in the non-doped structures are almost identical to those  
with electron doping, which confirmed that the Pb-site instability is from
increased Pb-O distance. 

 The self-force constant of Ti decreases, and the Ti-O1 IFC increases in the
 non-doped structure, just opposite to in the doped structures. This further
 shows that the increased occupation of the Ti $3d$ bands is the reason for the decreased
 Ti-site instability.

\begin{figure}[hbtp]
  \centering
  \includegraphics[width=0.43\textwidth]{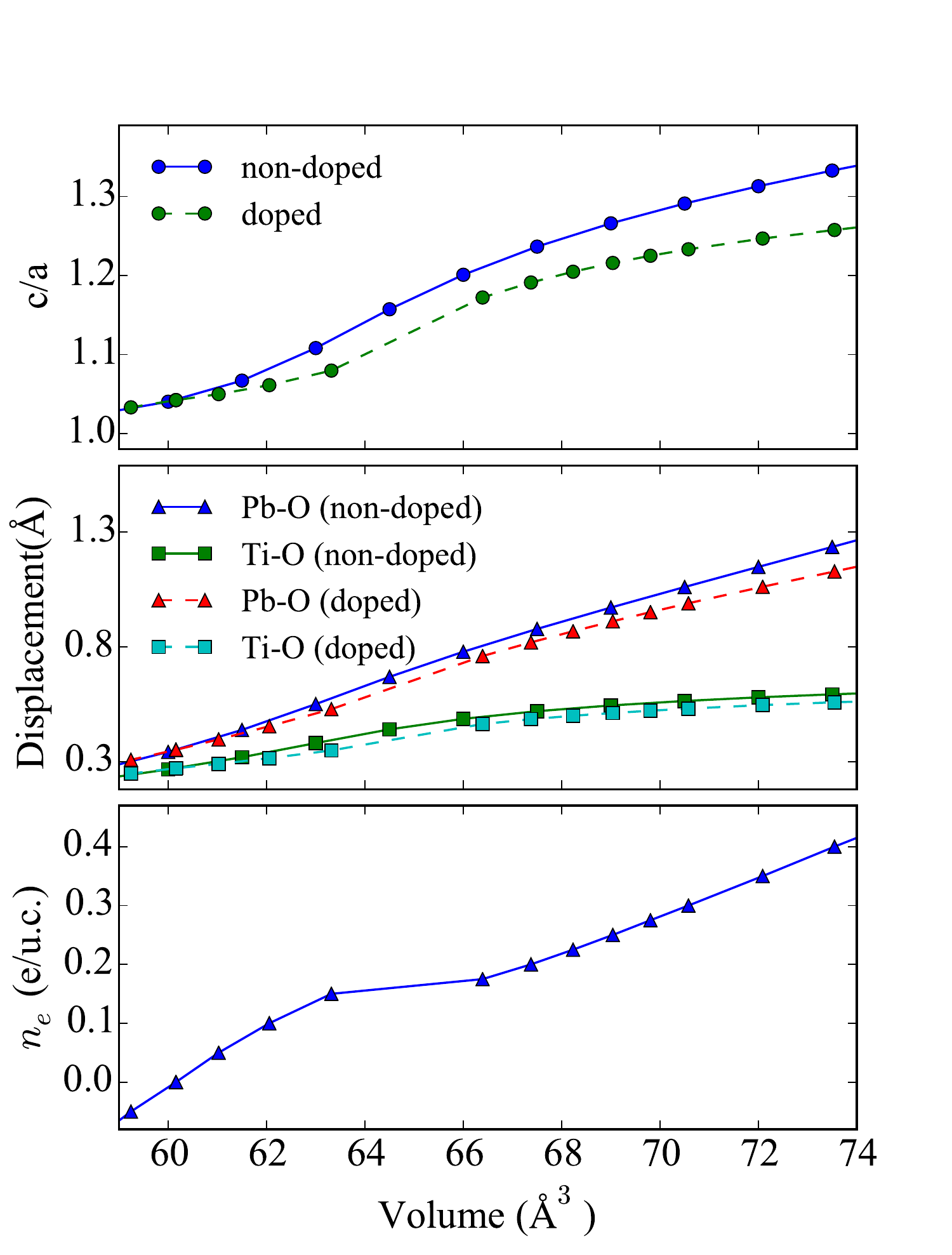}
  \caption{ (a) The ratio c/a as the function of lattice volume. (b) The
    relative displacements of Pb-O ($z(Pb)-z(O1)$) and Ti-O ($z(Ti)-z(O2)$) as functions of lattice volume. (c) The concentration of the doped electrons corresponding to the volume. The results for the non-doped structures were calculated by adding negative hydrostatic pressure.}
  \label{fig:varV}
\end{figure}

\begin{figure}[htbp]
  \centering
  \includegraphics[width=0.43\textwidth]{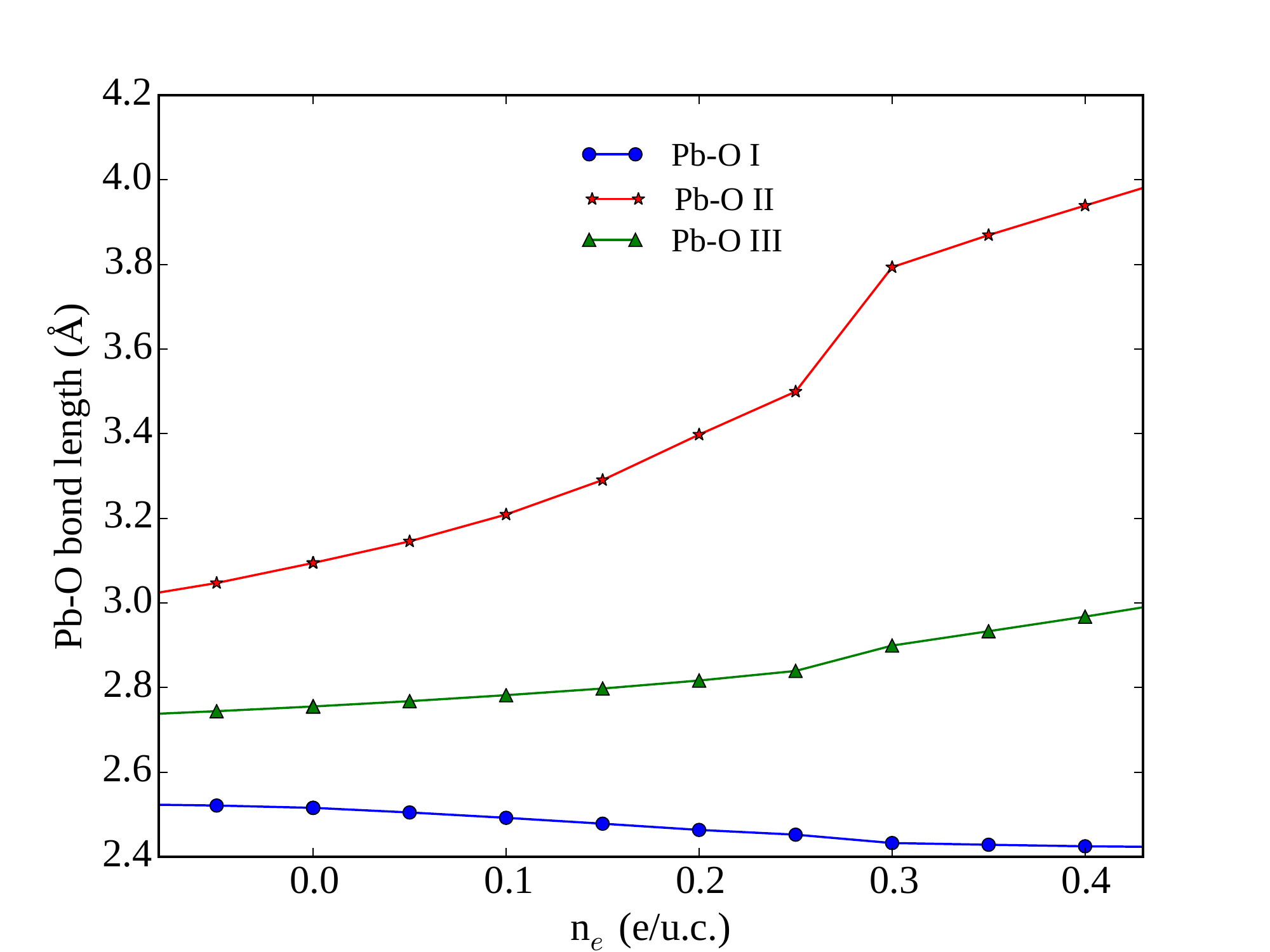}
  \caption{The change of lengths the Pb-O bonds with the concentration of doped electrons.\label{fig:bonds}}
\end{figure}

 We also noted that the change of the polar distortion with electron doping is
 very similar to that with the negative hydrostatic pressure in PbTiO$_3$. In
 both cases, the Pb-O bonds are stretched. We calculated the changes of the
 polar distortion of PbTiO$_3$ with negative hydrostatic pressure. The polar
 distortion increases with negative hydrostatic pressure in PbTiO$_3$, which
 agrees with previous studies\cite{PhysRevB.68.144105, 0953-8984-20-34-345207,
   wang2015negative}. The changes of the distortion with electron doping were
 plotted as functions of lattice volume and then compared to those with
 hydrostatic pressure, as shown in Fig.~\ref{fig:varV}. The changes of $c/a$ and
 cation-anion displacements with the same lattice volume are close in the two
 situations, indicating that the enhancement of the polar distortion might be from the same origin. It can be also seen that there is an anomalous enhancement of tetragonality and lattice volume as $n_e$ increases to about 0.3 e/u.c. (Fig.~\ref{fig:structvare}), which was also found in PbTiO$_3$ with negative pressure\cite{PhysRevB.68.144105}. The polar distortions in the electron-doped structure are smaller than those in the non-doped structures with the same volume, which may be the result of the screening effect.

  It is interesting to note that the changes of polar distortion with negative
  hydrostatic pressure and that with electron doping are opposite in BaTiO$_3$,
  where the polar distortion also increases with the negative hydrostatic
  pressure\cite{PhysRevB.68.144105, 0953-8984-20-34-345207}, but  decreases with
  the electron doping, though the volumes both increase in these two
  situations\cite{PhysRevB.68.144105,
    0953-8984-20-34-345207,wang2012ferroelectric}. The Ti-O bonds are stretched
  in the structure with negative hydrostatic pressure, therefore Ti becomes more under-bonded
, which requires a larger polar distortion.
  Whereas in the electron-doped structure, the Ti-O short-range repulsion is
  enhanced due to the electrons on the Ti $3d$ bands and the suppression of Ti PJTE, leading to the reduced
  Ti-site instability. In both cases, the Ba-O bonds are stretched, leading to
  the enhanced Ba-O instability, which is however not enough to result in a
  polar distortion in electron-doped BaTiO$_3$. Consequently, the polar
  distortion in BaTiO$_3$ decreases with electron doping. It can be seen that
  the change of polar distortion in BaTiO$_3$ and PbTiO$_3$ can be uniformly
  viewed as the result of the decreased of A-site instability and increased
  B-site instability.

 We examined the Pb-O bonds to see how the stretching of the Pb-O bonds affects
 the polar distortion. The lengths of the Pb-O bonds are plotted in Fig.
 \ref{fig:bonds}.  As the doped electrons
 enlarge the sizes of Ti ions and the lengths of Ti-O bonds, the Pb-O chain
 consisting of alternating type I and type II bonds are stretched. The changes
 of the type I bond lengths are relatively small, while those of the type II
 bond lengths are much larger. The reason is that the
 Pb-O electron hybridization stabilizes the short Pb-O type I bonds. And the
 enhanced Pb PJTE causes a larger displacement of Pb towards the side
 of type I bonds.

There are some similarities between the electron-doped PbTiO$_3$ and the NCSM
LiOsO$_3$. Firstly, the A-site instability drives the polar distortion.
Secondly, in both structures, the polar instabilities are due to the short-range
interactions, which is the PJTE (or equivalently, the covalent interaction) for PbTiO$_3$, and short-range
Coulomb interaction for LiOsO$_3$. Thirdly, the Fermi energy is in the B-site
bands and outside the energy range of the electronic states related to the A-O
interaction. In LiOsO$_3$, it was found that the ferroelectricity is due to the
Li-O displacement, whereas the Fermi level lies in the Os
bands~\cite{PhysRevB.89.201107,PhysRevB.90.094108,PhysRevB.90.195113,PhysRevB.91.064104}.
These similarities infer that the polar distortion caused by short-range interactions and the metallicity can
coexist if they are from different atoms. 

 The mechanism of the persistent or even the enhancement of the polar distortion
 in PbTiO$_3$ with electron doping presented in this paper should be
 transferable to other lone-pair driven ferroelectric materials, like in
 PbVO$_3$\cite{PhysRevB.73.094102}, BiFeO$_3$~\cite{PhysRevB.74.224412,PhysRevB.93.174110},
 SnTiO$_3$\cite{PhysRevB.91.035112}, BiMnO$_3$~\cite{doi:10.1021/cm010090m}. In
 these materials, the electronic states corresponding to the lone-pair mechanism are away from the Fermi energy if electrons are doped. Whereas the
 bottom of conduction bands in these materials are often the B-site states. Thus
 the doping of electrons can be seen as a selective enlargement of the B-site
 ion radius, which stretches the A-O bonds. Therefore, enhancement of polar
 distortion in these materials similar to that in PbTiO$_3$ is likely to emerge. These results also imply that the lone-pair
 stereoactive ions can be used as the A-site ions in perovskites to form NCSMs.
 By selecting a B-site element (or elements) with suitable ionic radius and
 itinerant electrons, lone-pair driven non-central-symmetric metal may be designed.
\section{Conclusion}
 In this work, we investigated the effect of electron doping on the lone-pair
driven polar distortion by carrying out density functional theory studies on
PbTiO$_3$. We found that the polar distortion is enhanced with electron doping
in PbTiO$_3$ even when the long-range Coulomb interaction is screened out by the doped
electrons. The analysis on the phonons and electronic
states show the mechanism for the persistent of the polar distortion: the
lone-pair mechanism, which is the driving force of the polar
distortion, is not strongly affected by the electron doping because the energy
range of the related electron states is far enough from the
Fermi energy. We also found that the enhancement of the polar distortion in
PbTiO$_3$ is due to the increasing of the Ti ion radius, which caused the
increasing of the
Pb-O distance. These results show that the
lone-pair driven polar distortion and the metallicity can coexist, and it is
highly expected that the lone-pair stereoactive ions can be used in designing NCSMs.

\begin{acknowledgments}
The work was supported by the National Basic Research Program of China (Nos. 2014CB921001 and 2012CB921403), the National Natural Science Foundation of China (No. 11134012), and the Strategic Priority Research Program (B) of the Chinese Academy of Sciences (No. XDB07030200).
\end{acknowledgments}
\bibliography{mybib}

\end{document}